\documentclass[aps,pra,twocolumn,groupedaddress,showpacs,10pt]{revtex4-1}
\usepackage{hyperref}
\usepackage{bm}
\usepackage{graphicx}

\begin{document}


\title{Appearance of Gibbs states in quantum-state tomography}


\author{Jochen Rau}
\email[]{jochen.rau@q-info.org}
\homepage[]{www.q-info.org}
\affiliation{Institute for Theoretical Physics,
University of Ulm,
Albert-Einstein-Allee 11,
89069 Ulm, 
Germany}
\affiliation{Department of Engineering,
RheinMain University of Applied Sciences,
Am Br\"uckweg 26,
65428 R\"usselsheim, 
Germany}
\thanks{permanent address}


\date{\today}

\begin{abstract}
I investigate the extent to which the description of quantum systems by Gibbs states can be justified purely on the basis of tomographic data, without recourse to theoretical concepts such as infinite ensembles, environments, information, or to the systems' dynamics.
I show that the use of Gibbs states amounts to a relevance hypothesis, which I spell out in detail.
This hypothesis can be subjected to statistical hypothesis testing and hence assessed on the basis of the experimental data.
\end{abstract}


\maketitle



\section{\label{intro}Introduction}

To describe the static or dynamic properties of a macroscopic quantum system,
typically only a few observables $\{G_a\}$ are deemed relevant -- 
for example, the system's constants of the motion (if static), slow observables (if dynamic), or
observables pertaining to some subsystem of interest.
In statistical mechanics the system is then assigned that quantum state 
which, while reproducing the observed values $\{g_a\}$ of the relevant observables,
maximizes the
\textit{von Neumann entropy} 
\begin{equation}
	S[\mu]:=-\mbox{tr}(\mu\ln\mu)
	;
\label{vonNeumann}
\end{equation}
i.e., its state -- which I shall denote by $\mu_g$ -- is determined  by the maximization
\begin{equation}
	\mu_g:=\arg \max_{\mu\in g} S[\mu]
	,
\end{equation}
where $\mu\in g$ is short for the constraints $\langle G_a\rangle_\mu=g_a \forall a$.
It has the \textit{Gibbs form}
\begin{equation}
	\mu_g 
	\propto
	\exp(-\lambda^a G_a)
\label{canonical}
\end{equation}
with Lagrange parameters $\{\lambda^a\}$ and a constant of proportionality (the inverse of the partition function) chosen to ensure state normalization, $\mbox{tr}\mu_g=1$.
For ease of notation I adopted here the Einstein convention that identical upper and lower indices are to be summed over.
In the special case that only the system's energy is relevant, the set $\{G_a\}$ contains just the Hamiltonian, and the associated Lagrange parameter is the inverse temperature;
the Gibbs state is then a canonical state.
While they first arose in the context of statistical mechanics, Gibbs states
nowadays play an important role also on smaller scales.
For instance, they have been employed successfully in nanoscale thermodynamics \cite{allahverdyan:nano,allahverdyan:work,janzing:molecular}, high energy physics \cite{becattini:thermal}, or 
incomplete quantum-state tomography \cite{buzek:spin,buzek:aop,buzek:reconstruction,buzek:jmo,buzek:lnp,buzek:motional,deleglise:cavity}.

Why entropy maximization, and hence the use of the Gibbs form, should be the proper paradigm for constructing the quantum state has been the subject of much debate.
The classic textbook argument in statistical mechanics relies on an idealization, the thermodynamic limit:
The system of interest is viewed as but one member of a fictitious infinite ensemble of identically prepared systems. 
If the global state of this fictitious ensemble is constrained by sharp values (not expectation values) for the totals of the relevant observables
then the reduced state of any single member of the ensemble has the Gibbs form  \cite{balian+balazs}.
Recent research suggests that one can do without such fictitious ensembles and derive the Gibbs form just as well directly  from a few generic assumptions, as long as the state in question pertains to a subsystem coupled to a sufficiently large environment \cite{PhysRevLett.96.050403,popescu:entanglement}.
Another popular argument invokes the intimate connection between entropy and information:
By maximizing entropy, Gibbs states discard to a maximal extent all information (and thus retain no spurious bias) as to irrelevant degrees of freedom;
so they carry  information solely about the relevant ones.
This insight is at the heart of the information-theoretic approach to statistical mechanics  \cite{jaynes:info1,jaynes:info2,katz:book,baierlein:book}.
Yet another line of reasoning, going back to Boltzmann's $H$ theorem \cite{boltzmann:h-theorem}, brings into play the system's effective dynamics on some coarse-grained level of description, in particular its tendency to increase entropy \cite{PhysRevLett.101.190403,PhysRevE.79.061103,1367-2630-13-5-053009,riera:thermalization}.
Such arguments rely on the existence of disparate time scales in the system  \cite{rau:physrep}.
Finally, some authors in both the statistics \cite{shore+johnson,skilling:axioms} and physics \cite{tikochinsky:consistent} communities have argued (for the classical case only) that the maximum entropy paradigm is mandated by logical consistency;
but this point of view remains controversial \cite{uffink:maxent,uffink:constraintrule}.

In the present paper I wish to add a different perspective.
State construction via maximum entropy is a special instance of a much broader task:
estimating a quantum state on the basis of imperfect data.
Experimental data are in fact {never} perfect, not even for simple systems, because the investigated samples have finite size, measurement devices have limited accuracy, and possibly -- as is the case in statistical mechanics -- the observables measured are not informationally complete.
So in practice, data \textit{never} specify a unique quantum state.
Rather, among the many states compatible with the data one must infer the most probable one,
using suitable statistical estimation techniques.
Such techniques have become an indispensable tool for data analysis in modern quantum physics experiments and  go by the name of \textit{quantum-state tomography} \cite{dariano:tomographic,1367-2630-15-12-125020}.

If the maximum entropy paradigm may be thus subsumed under the broader framework of quantum-state tomography then perhaps the latter can shed some light on the question as to when and how Gibbs states arise.
Exploring the extent to which this is indeed possible, is the purpose of the present paper.
Consequently, I shall tackle the issue of Gibbs states solely with the help of statistical methods from quantum-state tomography \textit{and nothing else;} 
in particular, without any recourse to the thermodynamic limit, environments, the concept of information, or dynamics.

The remainder of the present paper is organized as follows.
In Sec. \ref{tomography}, I will review some basic concepts of quantum-state tomography.
In Sec. \ref{sanov}, I will focus on the situation where the experimental data come in the form of sample means of some informationally incomplete set of observables.
I will show that in this case one can apply the quantum Sanov theorem to find the asymptotics of the pertinent likelihood function.
The subsequent Sec. \ref{relevance} is then crucial for the understanding of Gibbs states:
I will argue that the use of the Gibbs form is tantamount to a statistical ``relevance hypothesis'', for which I shall give a precise mathematical formulation. 
In Sec. \ref{modelselection}, I will discuss how the likelihood of this hypothesis and of possible rival hypotheses may be assessed in the light of experimental data.
Finally, in Sec. \ref{discussion}, I shall conclude with a brief summary and a few additional remarks.

\section{\label{tomography}Quantum-state tomography}

It is possible to know the precise state of an individual quantum system \textit{after} a measurement:
For instance, if measurement of some observable returns one of its non-degenerate eigenvalues then after the measurement, the system is known with certainty to be in the associated eigenstate.
(Precise knowledge of the post-measurement state hinges thus on precise knowledge of the measured observable.
Strictly speaking, the latter presupposes additional preceding measurements.)
Yet it is impossible to reconstruct, based on measurements on the individual system alone, its state \textit{before} the measurement \cite{hartle:individual}. 
Such a reconstruction requires instead measurements on many identically prepared copies; 
and even then, as the number of copies is always finite (let alone the limited accuracy of measurement devices), can never be perfect.
Thus in practice, measurement never yields a unique quantum state.  
Indeed, current experiments that implement quantum circuits or probe fundamental aspects of quantum information in many-body systems work with typical sample sizes of several hundreds or thousands, leading to statistical errors of up to $10$\% \cite{haeffner:scalable}.
Under these circumstances one can only aspire to identify among the many states compatible with the data the \textit{most probable} one.
This requires the use of suitable statistical estimation techniques and is the subject of \textit{quantum-state tomography} \cite{dariano:tomographic,1367-2630-15-12-125020}.

\begin{figure}[tbp]
\begin{center}
\includegraphics[width=8.5cm]{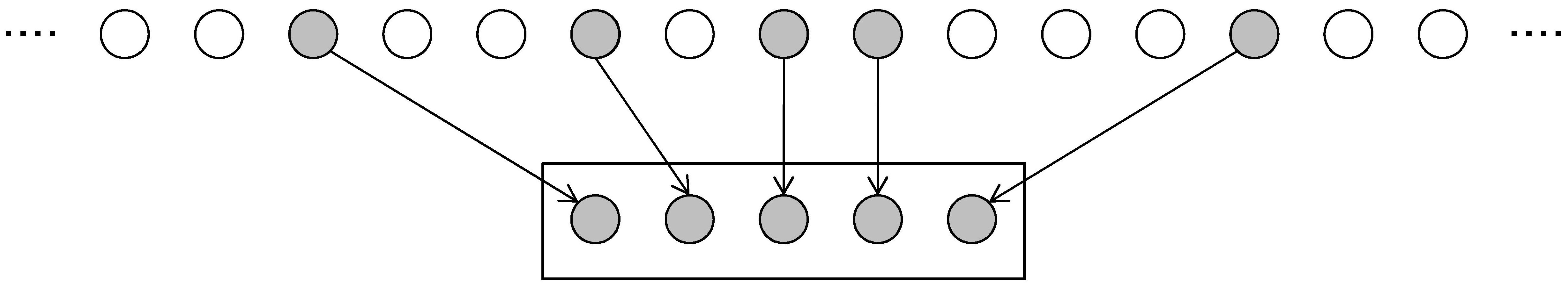}
\end{center}
\caption{Exchangeable sequence of quantum systems.
One obtains a finite exchangeable sequence of $L$ quantum systems by drawing randomly $L$ systems from a fictitious infinite symmetric sequence.
}
\label{exchangeable}
\end{figure}

Identical preparation of copies means that these form an \textit{exchangeable sequence} \cite{caves:definettistates}.
Such a sequence has finite length $L$, which may be chosen freely.
It can be thought of as drawn randomly from a fictitious infinite sequence of systems whose order is irrelevant  (Fig. \ref{exchangeable}).
Exchangeability entails two basic properties for the $L$-body state $\rho^{(L)}$ of the sequence:
(i) 
it is symmetric under permutation of the constituents;
and
(ii)
since the exchangeable sequence of length $L$ can always be considered a subsequence of a longer, equally exchangeable sequence of length $L+1$, 
the state $\rho^{(L)}$ can be written as a marginal of $\rho^{(L+1)}$.

Exchangeability is more than mere symmetry.
For example, the $2$-body density matrix $\rho_{AB}=|\psi_{AB}\rangle\langle\psi_{AB}|$
associated with the Bell state $|\psi_{AB}\rangle = (1/\sqrt{2}) (|00\rangle +|11\rangle)$
is invariant under permutation of the constituents and hence meets the symmetry criterion; 
yet it cannot be written as the marginal of an equally symmetric $3$-body state and thus fails to meet the second criterion for exchangeability.
Exchangeable is also not the same as independent and identically distributed (i.i.d.):
in general, it is $\rho^{(L)}\neq \rho^{\otimes L}$.
Rather, by a quantum generalization \cite{caves:definettistates} of the classical de Finetti theorem \cite{definetti:book}, 
the state of an exchangeable sequence can always be represented as an incoherent mixture of i.i.d. sequences $\rho^{\otimes L}$, 
\begin{equation}
	\rho^{(L)} = \int d\rho\ \mbox{prob}(\rho)\ \rho^{\otimes L}
	,
\label{definetti_state}
\end{equation}
with respective weights $\mbox{prob}(\rho)$,
where the integral is over all normalized single-particle states.
Conversely, any state of this form describes an exchangeable sequence.
The de Finetti representation shows that an exchangeable sequence may well exhibit classical correlations.
However, it never exhibits entanglement.

\begin{figure}[tbp]
\begin{center}
\includegraphics[width=8cm]{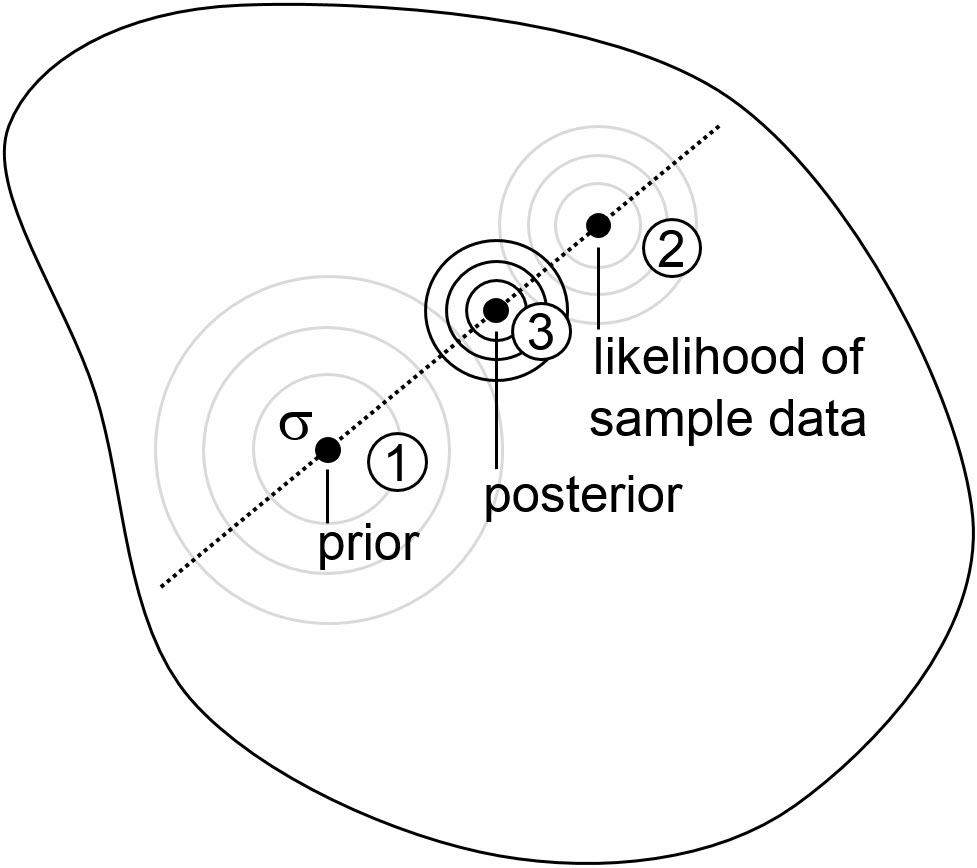}
\end{center}
\caption{Learning from sample data.
(1) Before the experiment an exchangeable sequence is characterized by some prior probability distribution $\mbox{prob}(\rho)$ in single-constituent state space.
If, say, on theoretical grounds one expects the members of the sequence to be in a state close to $\sigma$ then this prior will be peaked around $\sigma$. 
It has a finite width reflecting the finite degree of confidence in this prior bias.
(2)
Investigation of a sample yields data $D$.
Associated with these data is a likelihood function $\mbox{prob}(D|\rho^{\otimes N})$, typically peaked around some other state which might be close to, but is usually not equal to $\sigma$.
The likelihood function, too, has a finite width, reflecting the finite size $N$ of the sample (and possibly other sources of error).
(3)
According to Bayes' rule, multiplying the prior with the likelihood function yields the posterior $\mbox{prob}(\rho|D)$.
The latter is typically narrower than the prior, reflecting the growing confidence in the state estimate as experimental data accumulate.
The center of the posterior has shifted from the original bias $\sigma$ to a new state interpolating between $\sigma$ and the center of the likelihood function.
}
\label{learning}
\end{figure}

Exchangeable sequences are the ``raw material'' of quantum-state tomography.
The uncertainty about the state of an individual constituent is reflected in the density function $\mbox{prob}(\rho)$;
the latter may be considered (in somewhat loose terminology \cite{caves:definettistates}) the {probability distribution} for the unknown single-constituent state.
To learn more about this state, a sample of size $N$ ($N<L$) is taken from the exchangeable sequence and a measurement performed on it, yielding data $D$.
Afterwards the remaining $L-N$ systems (i.e., the original sequence minus the sample) still form an exchangeable sequence whose state has the above de Finetti representation;
yet the probability distribution featuring in this de Finetti representation must be updated according to a quantum generalization of \textit{Bayes' rule} \cite{schack:bayesrule},
\begin{equation}
	\mbox{prob}(\rho|D) \propto \mbox{prob}(D|\rho^{\otimes N}) \,\mbox{prob}(\rho)
	,
\label{bayesrule}
\end{equation}
where the probabilities denote (from left to right) the posterior, likelihood function, and prior, respectively, and the constant of proportionality is independent of $\rho$.
This Bayesian update encapsulates the process of \textit{learning} from sample data (Fig. \ref{learning}).

Upon investigation of additional samples, Bayes' rule is iterated, leading to consecutive updates of the posterior.
As more and more data accumulate -- by investigating more samples or increasing their sizes -- the posterior narrows until eventually its width falls below some desired error bound. 
Then within this error, the location of the posterior peak is the best estimate for the unknown quantum state.
In the hypothetical limit of infinite sample size, informationally complete measurements and perfectly accurate measurement devices the posterior converges towards the likelihood function, which in turn approaches a delta function.
The state estimate is then determined -- to perfect accuracy -- by experimental data only and becomes independent of the prior.
(It is the fact that this is possible, at least in principle, which gives operational meaning to the notion of ``state''.)
Against this backdrop many state estimation techniques focus from the outset on the likelihood function, equating the location of its peak with the most plausible state estimate;
such techniques fall into the class of \textit{maximum likelihood} methods \cite{hradil:estimation,james:qubits,hradil:lnp}.
In contrast, methods that take into account the residual influence of the prior (which in practice is always present and for small samples may be quite significant) are termed \textit{Bayesian} \cite{audenaert:kalman,rau:evidence,PhysRevA.84.012101}.

Strictly speaking, even in the above hypothetical limit the posterior coincides with the likelihood function only if the prior has support on the entire state space.
The prior reflects any theoretical constraint or bias that one may have, prior to measurement, as to the parametric form or parameter values of the quantum state.
As long as one knows nothing or little about the state \textit{a priori}, this prior is broad and  indeed has full support.
But as soon as one has advance knowledge that constrains the quantum state to some finite region or proper submanifold of state space, the prior has support on this region or submanifold only;
and so will the posterior, \textit{regardless of the data} \cite{fuchs:priors}. 
In Sec. \ref{relevance}, I shall argue that such \textit{a priori} restrictions play an important role in the understanding of Gibbs states.

\section{\label{sanov}Sanov likelihood}

\begin{figure}[tbp]
\begin{center}
\includegraphics[width=8cm]{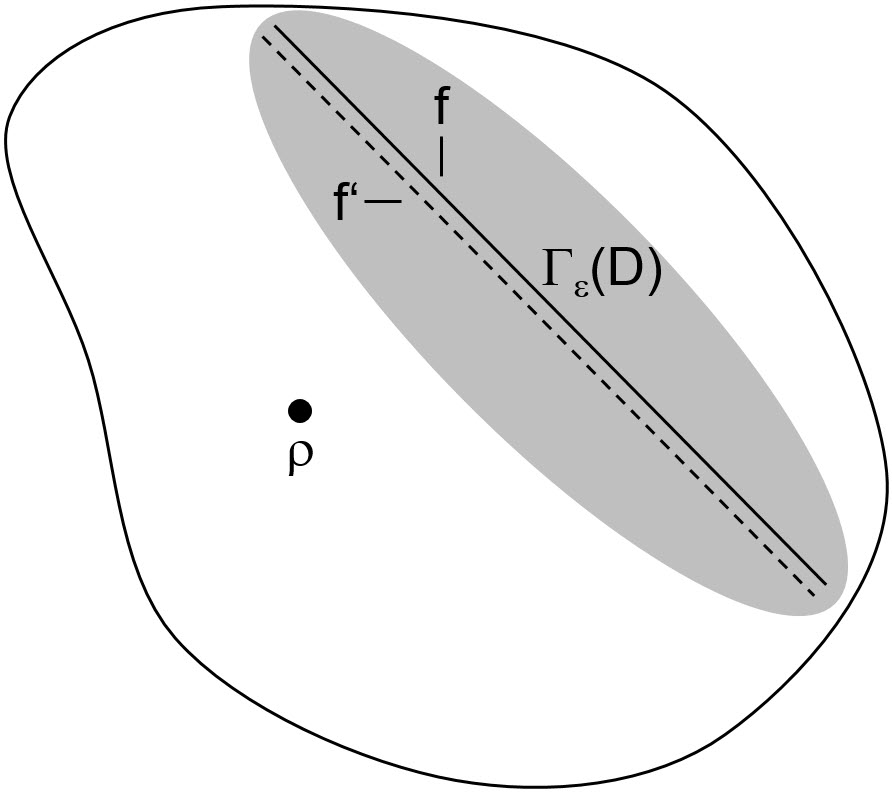}
\end{center}
\caption{Relating data to sample means.
The set $\Gamma_\epsilon(D)$ contains all states compatible (up to some error $\epsilon$) with observed data $D$ (shaded area).
In order to encompass sample means $\{f_b\}$ this set must contain all states yielding $\langle F_b\rangle=f_b \forall b$ (solid line).
In the above example, $\Gamma_\epsilon(D)$ is so large that it also encompasses  other values $\{f_b'\}$ for the sample means (dashed line).
}
\label{compatible}
\end{figure}

I suppose that the experimental data consist in a set of sample means $\{f_b\}$ gleaned from a sample of large but finite size $N$.
These sample means may have been obtained directly by measurement of the pertinent observables $\{F_b\}$, or inferred indirectly from other data $D$ ---
a possibility which is particularly relevant when the sample means pertain to observables that do not commute.
In the latter case I shall say that the observed data $D$ ``encompass'' sample means $\{f_b\}$ if and only if the set of states compatible with the data,
\begin{equation}
	\Gamma_\epsilon(D):=\left\{\mu\left|\mbox{prob}(D|\mu^{\otimes N})\geq 1-\epsilon\right.\right\}
\end{equation}
(up to some finite error parameter $\epsilon$, $0<\epsilon<1$, which is independent of sample size),
\textit{contains} the set of states yielding expectation values $\langle F_b\rangle_\mu=f_b \forall b$;
in short, $f\subseteq \Gamma_\epsilon(D)$.
When the set $\Gamma_\epsilon(D)$ is large, the data might encompass not just $\{f_b\}$ but also different values $\{f_b'\}$ for the sample means (Fig. \ref{compatible}).
On the other hand, if the set $\Gamma_\epsilon(D)$ contains only $f$ but no other $f'$, and is moreover the smallest set to do so, then I shall say that the data \textit{amount to} having measured the sample means $f$.
With this understanding, the likelihood of measuring sample means $f$ reads 
\begin{equation}
	\mbox{prob}_\epsilon(\{f_b\}|\rho^{\otimes N}):=\inf_D \left\{\left. \mbox{prob}(D|\rho^{\otimes N}) \right| {f\subseteq \Gamma_\epsilon(D)} \right\}
	.
\end{equation}

Defined in the above way, the likelihood generally depends on the error parameter $\epsilon$.
For large sample sizes $N$, however, the infimum on the right-hand side behaves asymptotically as
\begin{equation}
	\inf_D\, \{\ldots\} \sim \exp\left[-N \min_{\mu\in f} S(\mu\|\rho)\right]
\end{equation}
and hence loses its dependence on $\epsilon$;
this follows from the quantum generalization \cite{10.1088/0305-4470/35/50/307,10.1007/s00220-005-1426-2,10.1007/s00220-008-0417-5} of the classical Sanov theorem \cite{sanov:theorem,cover:book,deuschel:book}.
Here
\begin{equation}
	S(\mu\|\rho):=
	\left\{
\begin{array}{ll}
	\mbox{tr}(\mu\ln\mu - \mu\ln\rho) & : \mbox{supp}\;\mu\subseteq\mbox{supp}\;\rho \\
	+\infty & : \mbox{otherwise}
\end{array}
\right.
\label{relativeentropy}
\end{equation}
denotes the {relative entropy} of the two states $\mu$ and $\rho$ \cite{ochs:properties,wehrl:rmp,donald:cmp,vedral:rmp}.
In other words, for large $N$ the likelihood function behaves as
\begin{equation}
	\mbox{prob}(\{f_b\}|\rho^{\otimes N}) \sim \exp[-N S(\mu^\rho_f\|\rho)]
\label{likelihood}
\end{equation}
with 
\begin{equation}
	\mu^\rho_f := \arg \min_{\mu\in f} S(\mu\|\rho)
	,
\end{equation}
independently of $\epsilon$.
Due to its close connection to the quantum Sanov theorem, I shall call this asymptotic likelihood the \textit{Sanov likelihood}.
The state $\mu^\rho_f$ which, under given constraints on the expectation values $\{\langle F_b\rangle\}$, minimizes the relative entropy with respect to the ``reference state'' $\rho$ has the generalized Gibbs form \cite{ruskai:minrent}
\begin{equation}
	\mu^\rho_f
	\propto
	\exp[(\ln\rho-\langle\ln\rho\rangle_\rho)-\kappa^b F_b]
\label{generalized_gibbs}
\end{equation}
with Lagrange parameters $\{\kappa^b\}$ and the constant of proportionality again chosen to ensure $\mbox{tr} \mu^\rho_f=1$.

There is the special case where the sample means $\{f_b\}$ are informationally complete.
In this case the data determine a unique tomographic image (i.e., center of the likelihood function) $\mu$, the sole state to yield $\langle F_b\rangle_\mu=f_b \forall b$.
The quantum Sanov theorem then reduces to the quantum Stein lemma \cite{hiai+petz,ogawa+nagaoka,brandao+plenio},
and the asymptotic likelihood function becomes
\begin{equation}
	\mbox{prob}(\mu|\rho^{\otimes N}) \sim \exp[-N S(\mu\|\rho)]
	.
\label{stein_likelihood}
\end{equation}
I call this the \textit{Stein likelihood}.
Thanks to a mixing rule for the relative entropy \cite{PhysRevA.84.052101}, the Stein likelihood satisfies
\begin{eqnarray}
	\lefteqn{\mbox{prob}(\mu|\rho^{\otimes N})\cdot \mbox{prob}(\mu'|\rho^{\otimes N'}) 
	\propto}
	\nonumber \\
	&&
	\mbox{prob}(\textstyle\frac{N}{N+N'}\mu+\textstyle\frac{N'}{N+N'}\mu'|\rho^{\otimes (N+N')})
\end{eqnarray}
with a constant of proportionality that does not depend on the state $\rho$.
So for the purposes of Bayesian updating via Eq. (\ref{bayesrule}), obtaining first a tomographic image $\mu$ from a sample of size $N$ and subsequently a tomographic image $\mu'$ from another sample of size $N'$ is tantamount to obtaining the weighted average of $\mu$ and $\mu'$ from the combined sample of size $N+N'$;
sequential or joint processing of the data both yield the same posterior. 
In other words, in the asymptotic limit considered here it does not matter how the system copies under investigation are grouped into samples.

\section{\label{relevance}Relevance hypothesis}

The statement: ``The observables $\{G_a\}$ are relevant'' 
entails two distinct assertions:
(i)
The expectation values of the $\{G_a\}$ completely determine the state estimate (and all predictions following from it);
and 
(ii)
any update of this state estimate is determined by additional data pertaining to the $\{G_a\}$ only, and not by any other data.
(This notion of ``relevance'' is similar to the notion of ``consistency'' invoked -- in the classical case and for one special set of observables only -- in Ref. \cite{tikochinsky:consistent}.)
In this section, I shall prove that the relevance hypothesis imposes on the state estimate the Gibbs form (\ref{canonical}).

The first assertion implies that there must exist an algorithm $\{g_a\}\mapsto \rho$ assigning to any set of expectation values $\{g_a\}\equiv\{\langle G_a\rangle\}$ a unique state $\rho$.
This algorithm need not necessarily be the maximum entropy algorithm.
More generally, when the expectation values $\{g_a\}$ are not known exactly but only their probability distribution $\mbox{prob}(g)$
then there must exist an algorithm assigning to this probability distribution a unique probability distribution of states,
$\mbox{prob}(g)\mapsto \mbox{prob}(\rho)$.

I will now show that the second, logically independent assertion singles out the maximum entropy algorithm, and hence the Gibbs form (\ref{canonical}).
In my proof I will invoke 
various states and sets of states which are illustrated in Fig. \ref{relevance}.
Let a tomographic measurement on a sample of large but finite size $N$ come in two versions, one informationally complete and the other informationally incomplete.
Whereas the complete version returns a unique tomographic image $\mu$,
the incomplete version merely returns sample means $\{g_a\}$ for the observables $\{G_a\}$.
The latter are consistent with the former, $g_a=\langle G_a\rangle_\mu$.
If indeed the observables $\{G_a\}$ are the relevant ones then by the second assertion, it must not make a difference which of the two data sets, complete or incomplete, is processed in the Bayesian update (\ref{bayesrule}).
Both must yield the same posterior, and so it must hold that
\begin{equation}
	\mbox{prob}(\mu|\rho^{\otimes N})\,\mbox{prob}(\rho)
	\propto
	\mbox{prob}(\{g_a\}|\rho^{\otimes N})\,\mbox{prob}(\rho)
\end{equation}
with a constant of proportionality that does not depend on $\rho$.
This requirement can only be met if either the prior $\mbox{prob}(\rho)$ vanishes
or else, by Eqs. (\ref{likelihood}) (with $f=g$) and (\ref{stein_likelihood}), the difference of relative entropies
$[S(\mu\|\rho)-S(\mu^\rho_g\|\rho)]$ is independent of $\rho$.
By the law of Pythagoras for the relative entropy \cite{petz:book}, this difference is itself a relative entropy,
\begin{equation}
	S(\mu\|\mu^\rho_g)
	=
	S(\mu\|\rho)-S(\mu^\rho_g\|\rho)
	.
\end{equation}
It ought to have the same value for all $\rho$ that have a non-vanishing prior probability.
Let $\sigma$ be one specific such state with non-vanishing prior probability.
Then for all other $\rho$, it must hold that
\begin{equation}
	S(\mu\|\mu^\rho_g) = S(\mu\|\mu^\sigma_g)
	\quad\forall\,
	\mu,\rho:\,\mbox{prob}(\rho)\neq 0
	.
\end{equation}
In the special case $\mu=\rho$ it is $g_a=\langle G_a\rangle_\rho=:g_a(\rho)$ and hence also $\mu^\rho_g=\rho$, so the left-hand side vanishes.
Then so must the right-hand side, and therefore
\begin{equation}
	\rho=\mu^\sigma_{g(\rho)}
	\quad\forall\,
	\rho:\,\mbox{prob}(\rho)\neq 0
	.
\end{equation}
Regardless of the experimental data, the admissible states ($\mbox{prob}(\rho)\neq 0$) are restricted \textit{a priori} to Gibbs states of the generalized form (\ref{generalized_gibbs}), with reference state $\sigma$ and $\{F_b\}=\{G_a\}$.

\begin{figure}[tbp]
\begin{center}
\includegraphics[width=8cm]{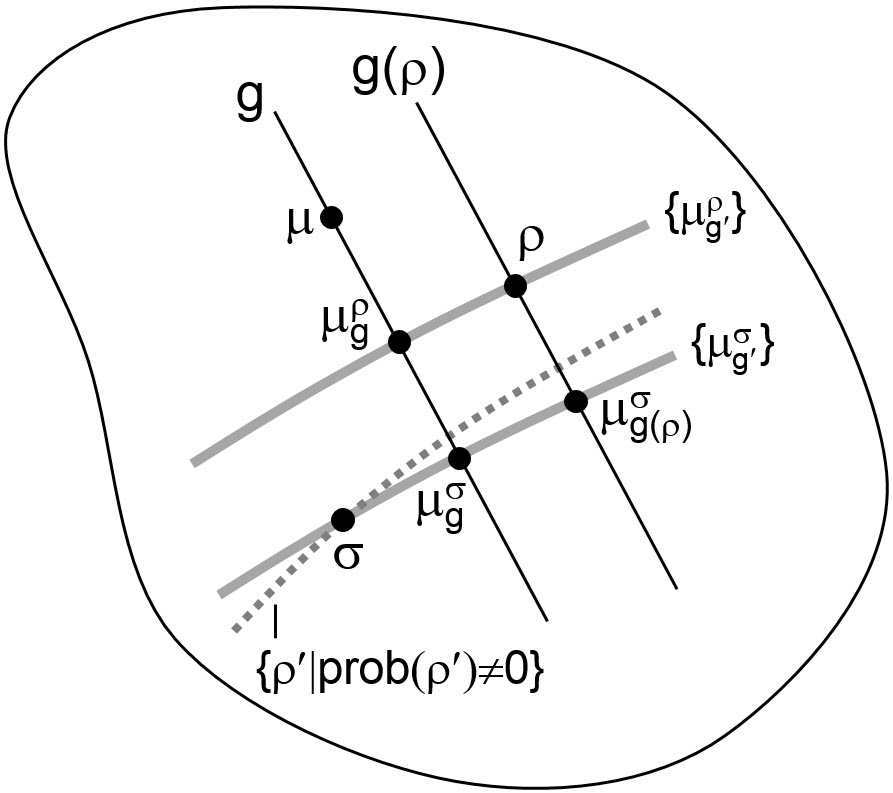}
\end{center}
\caption{States (dots) and sets of states (lines) featuring in the discussion of the relevance hypothesis.
A tomographic measurement on a large but finite sample comes in two versions, one complete and the other incomplete.
The complete version returns a unique tomographic image $\mu$;
whereas the incomplete version merely returns sample means $g$ for the informationally incomplete set of observables $\{G_a\}$.
The state $\mu$ is one out of the many states yielding $\langle G_a\rangle=g_a$ (left black line).
An arbitrary state $\rho$ yields expectation values $g(\rho)$,
as do all other states that lie on the right black line.
States which minimize the relative entropy with respect to $\rho$, while satisfying given constraints on the expectation values $\{\langle G_a\rangle\}$, form a proper submanifold of state space (upper grey line);
for $\langle G_a\rangle=g_a$, the pertinent state is $\mu^\rho_{g}$.
According to Bayes' rule, the posterior state estimate depends not only on the data but also on the prior. 
In particular, any estimate must be among the states with a non-vanishing prior probability (dotted line).
Let the latter include some specific state $\sigma$.
States which minimize the relative entropy with respect to this $\sigma$, while satisfying constraints on the $\{\langle G_a\rangle\}$, form another submanifold (lower grey line);
for $\langle G_a\rangle=g_a$ and $\langle G_a\rangle=g_a(\rho)$ the pertinent states are $\mu^\sigma_{g}$ and $\mu^\sigma_{g(\rho)}$, respectively.
If the relevance hypothesis holds then this last submanifold contains all states with non-vanishing prior probability;
so the lower grey line in fact covers the dotted line.
}
\label{fig_relevance}
\end{figure}

The reference state $\sigma$ may be any state that has a non-vanishing prior probability.
Among the states with non-vanishing prior probability there is usually (albeit not necessarily always) the totally mixed state.
If so, it will be most convenient to choose $\sigma$ to be the totally mixed state.
With the totally mixed state as the reference state, minimizing the relative entropy becomes equivalent to maximizing the ordinary von Neumann entropy; 
and then the generalized Gibbs form (\ref{generalized_gibbs}) reduces to the ordinary Gibbs form (\ref{canonical}), Q.E.D.

The relevance hypothesis has significant implications for quantum-state tomography.
It affects both the location (in state space) and the accuracy of the posterior state estimate:
Given the same data, different choices for the set of relevant observables generally lead to different state estimates with different degrees of confidence.
This can be illustrated with the simple example of state tomography on an exchangeable sequence of qubits.
The state space of a single qubit is the Bloch sphere, with the totally mixed state at its origin.
Let an incomplete tomographic measurement probe the Pauli spin component $X$, yielding sample mean $\bar{x}$.
Associated with this data is a likelihood function on the Bloch sphere.
It is broad on the two-dimensional plane containing states that yield $\langle X\rangle=\bar{x}$
and narrowly peaked in the direction perpendicular to this plane,
the latter width decreasing with increasing sample size.
Considering solely this likelihood function would lead to a maximum-likelihood state estimate equal or close to $\rho_{\rm ml}=(1+\bar{x}X)/2$.
However, as I discussed in Sec. \ref{tomography}, one must take into account also the prior;
and in particular, whether the prior has support on the entire Bloch sphere or on some subspace only.

\begin{figure}[tbp]
\begin{center}
\includegraphics[width=8cm]{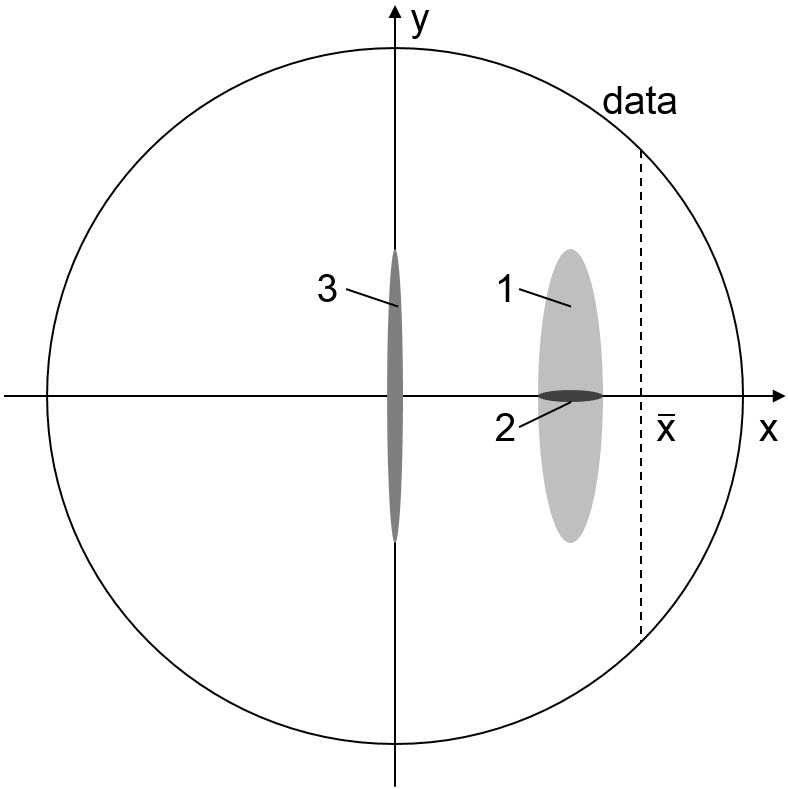}
\end{center}
\caption{Two-dimensional section ($z=0$) of the Bloch sphere.
The dashed line at $x=\bar{x}$ indicates states yielding $\langle X\rangle=\bar{x}$, the observed sample mean.
The shaded areas indicate location and width of the posterior when the relevant observables comprise
(1) all observables;
(2) only $X$;
or
(3) only $Y$.
In the first two cases the resultant state estimate (center of the posterior) lies somewhere between initial bias (totally mixed state) and data,
the precise location depending on the size of the sample.
The two cases differ in the degree of confidence as to the unmeasured observable $Y$.
In the third case the posterior equals the prior because the data carry no information about the then relevant observable $Y$.
}
\label{fig_blochsphere}
\end{figure}

For the present example I shall consider three cases:
(i)
all observables are relevant in the sense defined above;
(ii)
only $X$ is relevant (say, because the physical qubit is a spin in a ferromagnet which is strongly anisotropic in the $x$ direction);
or
(iii)
only $Y$ is relevant.
Whereas in the first case the prior has support on the entire Bloch sphere, in the other two cases it has support only on the $x$ or $y$ axis, respectively.
On the respective support, in the absence of further information, the prior is some broad symmetric distribution around the origin of the Bloch sphere.
Multiplying the respective priors with the likelihood function yields the respective posteriors.
These posteriors vary strongly from case to case;
they are depicted schematically in Fig. \ref{fig_blochsphere}.

\section{\label{modelselection}Model selection}

In the preceding section I have shown how the relevance hypothesis constrains state estimates \textit{a priori} to the Gibbs form (\ref{canonical}),
and how this affects quantum-state estimation.
One may wonder what in turn justifies the relevance hypothesis, and to which set of observables it should apply.

First of all, it is important to note that the relevant observables do not necessarily coincide with the observables which are being measured in an experiment;
an observable is not relevant simply because one happens to measure it.
In the example considered above the observable $X$ was measured.
And if the physical qubit was a spin in a strongly anisotropic ferromagnet with preferred direction along the $x$ axis then indeed, the observable $X$ would also be the relevant one.
But if the preferred direction of the ferromagnet was along the $y$ axis, the relevant observable would be $Y$ rather than $X$ --- 
even though $X$ was measured.
Rather than the experimental setup, the relevance hypothesis reflects prior knowledge about the \textit{physics} of the system.
As is familiar from statistical mechanics, the choice of relevant observables is usually linked to time scales:
If the system is in equilibrium then the relevant observables comprise the system's constants of the motion;
or else, provided the system's degrees of freedom evolve on two or more disparate time scales, the slow observables \cite{rau:physrep}.

The above warning notwithstanding, identifying measured with relevant observables is precisely what is being done -- implicitly -- in maximum-entropy quantum state estimation \cite{buzek:spin,buzek:aop,buzek:reconstruction,buzek:jmo,buzek:lnp,buzek:motional,deleglise:cavity}.
This shortcut (which is usually not spelt out as such) actually often works because for an observable to be measurable in practice, it must vary slowly;
and as long as the system exhibits a clear hierarchy of time scales, ``slow'' means indeed ``relevant''.

Yet when one investigates a hitherto unknown substance, a hierarchy of time scales or other clues as to the relevant observables are not available \textit{a priori;}
nor is there any assurance that the shortcut ``measured = relevant'' is warranted.
In such a situation one can only formulate conjectures as to the set of relevant observables.
The relevance hypothesis then becomes truly a \textit{hypothesis} in the statistical sense,
subject to experimental scrutiny and possibly refutation.
There might be several competing hypotheses, perhaps even including the hypothesis that the system cannot be described by Gibbs states at all --- that is, unless the set of relevant observables is extended to become informationally complete, in which case \textit{all} observables would be relevant.
Every proposal as to the set of relevant observables constitutes a \textit{statistical model,} in the sense that state estimates are constrained to a certain parametric form (Gibbs form) with a certain number of adjustable parameters (Lagrange parameters).
Choosing among rival proposals in the light of the experimental data then becomes an instance of statistical \textit{model selection}.
It is this scenario on which I shall focus in the present section.

First, some general remarks about statistical model selection may be in order.
In general, rival statistical models for the same experimental data differ in the number and type of adjustable parameters.
On the one hand, a model ought to be in good agreement with the data, which is best achieved with a large number of adjustable parameters;
yet on the other hand, excessive complexity must be avoided (``Occam's razor'').
The purpose of model selection is to render this trade-off quantitative. 
How this works in practice can be illustrated with a simple textbook example \cite{sivia:modelselection,sivia:book}.
Let $A$ be a simple model without adjustable parameter and $B$ a more complex model with one adjustable parameter $\lambda$. 
Which model is to be preferred on the basis of data $D$ will be determined by the ratio of their respective posterior probabilities.
Due to Bayes' rule, this ratio is given by 
\begin{equation}
	\frac{\mbox{prob}(A|D)}{\mbox{prob}(B|D)}
	=
	\frac{\mbox{prob}(D|A)}{\mbox{prob}(D|B)}
	\,
	\frac{\mbox{prob}(A)}{\mbox{prob}(B)}
	.
\end{equation}
As long as one does not have any strong \textit{a priori} preference for either of the models the right-hand side will be dominated by the first factor, the ratio of likelihoods.
By the law of total probability, the likelihood function of model $B$ reads
\begin{equation}
	\mbox{prob}(D|B)
	=
	\int d\lambda\ \mbox{prob}(D|\lambda,B)\,\mbox{prob}(\lambda|B)
	.
\end{equation}
Let $\lambda_0$ be that value of the adjustable parameter which yields the best fit with the experimental data.
Then the first factor in the integral, considered as a function of $\lambda$, will be peaked around a maximum at $\lambda_0$;
a typical shape is a Gaussian
\begin{equation}
	\mbox{prob}(D|\lambda,B)
	=
	\mbox{prob}(D|\lambda_0,B)
	\,
	\exp\left[-\frac{(\lambda-\lambda_0)^2}{2 (\delta\lambda)^2}\right]
\end{equation}
of some width $\delta\lambda$.
This width indicates the accuracy to which the parameter $\lambda$ is known after processing the experimental data. 
In contrast, the second factor in the integral describes the distribution of $\lambda$ \textit{prior} to processing the data;
this \textit{a priori} distribution has a larger width $\Delta\lambda>\delta\lambda$.
Provided the best fit $\lambda_0$ lies within the \textit{a priori} expected range, the ratio of likelihoods will scale as 
\begin{equation}
	\frac{\mbox{prob}(D|A)}{\mbox{prob}(D|B)}
	\sim
	\frac{\mbox{prob}(D|A)}{\mbox{prob}(D|\lambda_0,B)}
	\,
	\frac{\Delta\lambda}{\delta\lambda}
	.
\end{equation}
The first ratio on the right-hand side is typically smaller than one, favoring the more complex model $B$ because thanks to its adjustable parameter, $B$ can achieve a better fit with the data.
In contrast, the second ratio $(\Delta\lambda/\delta\lambda)$ is larger than one, favoring the simpler model $A$;
this is the quantitative manifestation of Occam's razor.
It is thus the relative size of these two ratios which will tip the balance in favor of one specific model.

The same logic applies to the identification of the most plausible set of relevant observables on the basis of experimental data.
Details of the pertinent statistical analysis have been spelt out by the author (in a different context) in previous publications for two basic scenarios.
In the first scenario \cite{PhysRevA.84.012101}, different hypotheses as to the set of relevant observables are formulated \textit{prior} to experiment, and subsequently the experiment is performed on a single sample only.
In the second scenario \cite{PhysRevA.84.052101}, which resembles more closely the way an unknown substance is investigated in practice,
measurements are performed first, before formulating any hypotheses.
Moreover, measurements are performed not just on a single sample but on multiple samples of the same unknown substance.
These samples need not -- in fact, ought not -- be in the same state.
Yet it is hypothesized that for all these samples the \textit{same} set of observables is relevant.
Their respective states should therefore all lie in the same submanifold of  Gibbs states, possibly with varying values for the associated Lagrange parameters.

For instance, the samples might all have been brought into contact with heat baths at different temperatures.
Then one hypothesis might say that they are now all in canonical states, the Hamiltonian (identical for all samples) being the sole relevant observable, with only the temperature varying across samples.
Another hypothesis might say that beyond the Hamiltonian there are further constants of the motion (again identical for all samples) which need to be added to the set of relevant observables, increasing the number of adjustable Lagrange parameters.
And a third, extreme hypothesis might claim that the samples have not thermalized fully, and  that hence a Gibbs form is not justified and \textit{all} observables remain relevant.
As in the textbook example above, increasing thus the number of adjustable parameters will lead to a successively better fit with the experimental data, yet at the expense of simplicity.
Again, goodness-of-fit has to be traded off against simplicity in a quantitative fashion.
The optimal trade-off yields the most plausible set of relevant observables.

\begin{figure}[tbp]
\begin{center}
\includegraphics[width=8cm]{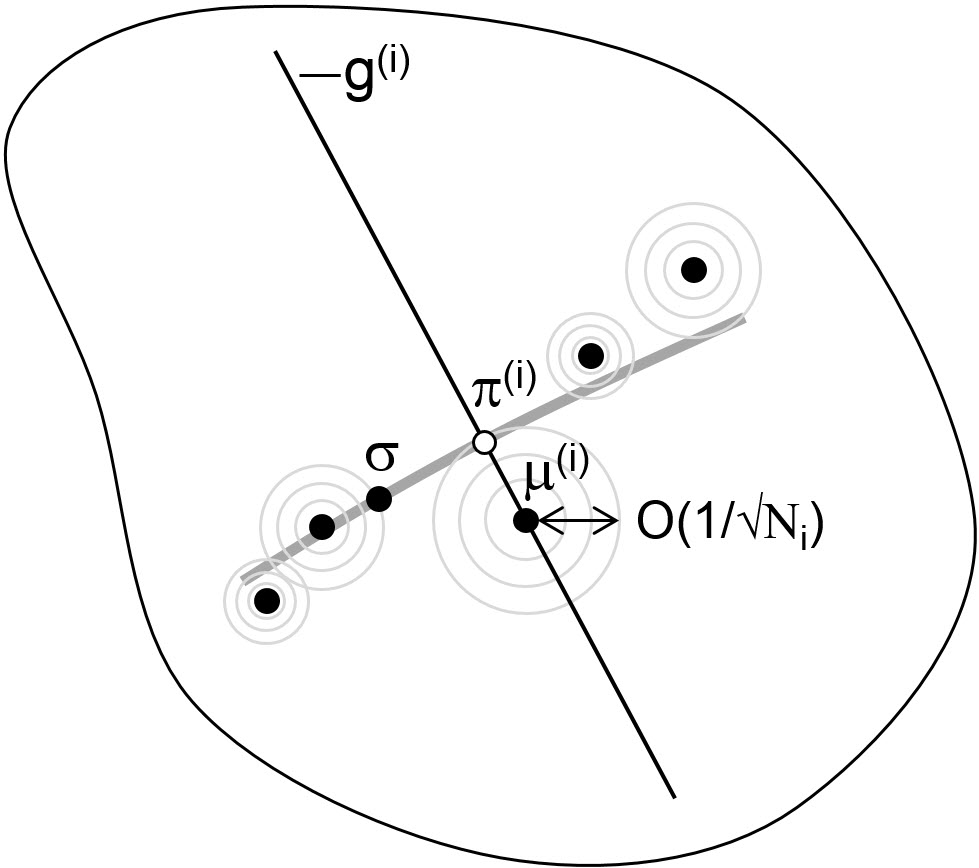}
\end{center}
\caption{Optimizing the set of relevant observables.
Black dots represent tomographic images for various samples of the same substance.
The concentric circles around them symbolize the associated likelihood functions, whose widths are an indicator of the measurement uncertainties.
These uncertainties vary by sample and typically scale as $1/\sqrt{N_i}$.
The relevance hypothesis, which is to be tested, claims that all data can be modelled on a joint manifold of Gibbs states (grey line) with reference state $\sigma$.
If so, the Gibbs state closest to a tomographic image $\mu^{(i)}$ is the state $\pi^{(i)}$ (open dot).
Both states yield for the relevant observables the same expectation values $\{g^{(i)}_a\}$,
i.e., they both belong to the set of states yielding $\langle G_a\rangle=g^{(i)}_a$ (black line).
}
\label{assess}
\end{figure}

Here I briefly summarize the key findings of the quantitative analysis;
details have been reported in Ref. \cite{PhysRevA.84.052101}.
Various differently prepared samples are subjected to the same tomography which, if not complete, must encompass at least all candidates for relevancy.
Let the $i$-th sample have large but finite size $N_i$,
and let tomography on this sample render the tomographic image $\mu^{(i)}$.
(In case the tomography is not complete, the image $\mu^{(i)}$ is constructed via ordinary maximum-entropy state estimation.)
Assuming that the finite sample size is the principal source of noise, error bars on the measurements are of the order $1/\sqrt{N_i}$.
As before, let $\sigma$ be a reference state (usually the totally mixed state) with non-vanishing prior probability,
and let ${\cal G}$ denote the hypothesis that for {all} samples the $p$ observables $\{G_a\}_{a=1}^p$ are the relevant ones in the sense defined above.
Given this hypothesis, states are constrained \textit{a priori} to Gibbs states of the generalized form $\mu^\sigma_g$.
For a tomographic image $\mu^{(i)}$ the closest such state is the state $\pi^{(i)}:=\mu^\sigma_{g^{(i)}}$, where $g_a^{(i)}:=\langle G_a\rangle_{\mu^{(i)}}$ (Fig. \ref{assess}).
In terms of these variables, and under certain reasonable additional assumptions spelt out in Ref. \cite{PhysRevA.84.052101}, the log-likelihood of the hypothesis ${\cal G}$ behaves asymptotically as
\begin{equation}
	\ln\mbox{prob}(D|{\cal G}) 
	\sim
	-\sum_i N_i S(\mu^{(i)}\|\pi^{(i)}) - \frac{p}{2}\sum_i \ln N_i
\end{equation}
modulo additive constants that do not depend on the choice of relevant observables.
Here $D$ is short for the totality of experimental data.
As long as there is no strong \textit{a priori} preference for a specific set of relevant observables, the difference of log-likelihoods of rival hypotheses dominates their relative posterior probabilities.

The above formula for the log-likelihood reflects in a quantitative fashion the expected trade-off between goodness-of-fit and simplicity.
On the right-hand side there are two contributions, both with a negative sign and thus ``penalizing'' -- in terms of likelihood -- the hypothesis ${\cal G}$.
The first contribution penalizes a bad fit to the data:
The further the Gibbs states $\pi^{(i)}$ are away from the original tomographic images $\mu^{(i)}$, the larger the relative entropies $S(\mu^{(i)}\|\pi^{(i)})$, and hence the larger the penalty.
To avoid this penalty, the set of relevant observables should be chosen sufficiently large.
In contrast, the second contribution embodies Occam's razor, penalizing excessive complexity:
The larger the number $p$ of relevant observables, the larger the penalty.
In order to avoid the latter penalty, the set of relevant observables should be kept as small as possible.
So in line with our general considerations, one must trade off these two penalties in order to find the most likely set of relevant observables.

In sum, in the absence of any prior knowledge about time scales or other clues as to the set of relevant observables, the most likely set can be determined via the above statistical analysis.
The use of the associated Gibbs form is then corroborated by the tomographic data only.

\section{\label{discussion}Conclusions}

In the preceding sections I explored the extent to which the use of Gibbs states can be understood with the help of methods from quantum-state tomography.
These methods apply to systems that are finite and isolated;
hence they require neither the limit $N\to\infty$ nor auxiliary concepts such as infinite ensembles or large environments.
(However, I did assume that sample sizes are large enough to justify the use of the asymptotic Sanov likelihood to good accuracy.)
Moreover, they refrain from invoking information-theoretical arguments or exploiting the system's dynamics.

Without any knowledge of the system's dynamics or other clues as to the relevant observables, of course, it is impossible to derive a Gibbs form from first principles.
Rather, in this situation the Gibbs form constitutes a statistical \textit{hypothesis} that can be supported or refuted only by data.
In Sec. \ref{relevance}, I gave a precise formulation of the relevance hypothesis which is behind the use of the Gibbs form.
And in Sec. \ref{modelselection}, I outlined the statistical tools needed to test this hypothesis in the light of experimental data.

Taken together, the statistical methods discussed in this paper comprise a toolbox which can be used to ascertain
(i)
whether a hitherto unknown substance, of which in particular the dynamics and the constants of the motion are not known, can be described by Gibbs states at all;
and if so,
(ii)
which observables are most likely the relevant ones.
The pertinent statistical analysis is based entirely on the tomographic data gleaned from a collection of differently prepared, finite samples.
It focuses on the relative likelihoods rather than posterior probabilities of rival hypotheses.
The former are a good proxy for the latter as long as there is no prior knowledge, and hence no \textit{a priori} bias in favor of any particular hypothesis.

Once the set of relevant observables, valid for the entire collection of samples, has been established, there remains the statistical task of estimating the values of the pertinent Lagrange parameters for any given \textit{individual} sample.
This is an interesting subject in itself, which has been dealt with elsewhere \cite{rau:evidence,PhysRevA.84.012101}.

%



%

\end{document}